\documentclass[modern]{aastex62}

\received{}
\revised{}
\accepted{}
\submitjournal{ApJ}

\shorttitle{A unified framework for X-shaped radio galaxies}
\shortauthors{Garofalo et al.}

\begin{document}

\title{A unified framework for X-shaped radio galaxies}

\correspondingauthor{David Garofalo}
\email{dgarofal@kennesaw.edu}

\author{David Garofalo}
\affil{Department of Physics, Kennesaw State University, Marietta GA 30060, USA}

\author{Ravi Joshi}
\affil{Kavli Institute for Astronomy and Astrophysics, Peking University, Beijing, 100871, China}

\author{Xiaolong Yang}
\affil{Kavli Institute for Astronomy and Astrophysics, Peking University, Beijing, 100871, China}

\author{Chandra B. Singh}
\affiliation{The Raymond and Beverly Sackler School of Physics and Astronomy, Tel Aviv University, Tel Aviv 69978, Israel}
\affiliation{South-Western Institute for Astronomy Research, Yunnan University, University Town, Chenggong, Kunming 650500, P. R. China}

\author{Max North}
\affil{Department of Information Systems, Kennesaw State University, Marietta GA 30060, USA}

\author{Matthew Hopkins}
\affil{Department of Physics, Kennesaw State University, Marietta GA 30060, USA}



\begin{abstract}
We propose a radically different picture for X-shaped radio galaxies compared to existing models as transition objects 
between cold mode accreting low spinning retrograde black holes and low spinning prograde black holes.  The model 
explains their smaller average black hole masses, their general aversion for cluster compared to isolated environments, 
the negligible difference in spectral index between primary and secondary jets despite a time difference in their formation, 
their absence among the most powerful radio quasars and radio galaxies, and their connection to the elusive FRI quasar 
class, among others. The key to their formation is cold gas accretion onto Schwarzschild black holes. 
\end{abstract}

\keywords{galaxies: active -- galaxies: radio galaxies -- galaxies: supermassive black holes}


\section{Introduction} 

X-shaped radio galaxies (XRG) are active galaxies (AGN) with two misaligned jets emerging from the central core constituting 
about $7\%$ of AGN with jets \citep{che07}. They have been known to exist for about half a century but have been difficult to 
connect to the general AGN phenomena. Models that explain some of the observations of XRG involve backflow \citep{cap02, hod11, ros17}
, jet precession \citep{den02, cap06}, spin-flip \citep{mer02, gop03} and dual AGN \citep{lal19}.  Observationally, the primary jet appears to 
be mostly of the FRII type \citep{lea92, sar18, jos19} although some FRI morphologies 
for the primary jet exist \citep{lea84, jon92, mur01}. Their luminosities lie close to 
the Owen-Ledlow boundary \citep{den02, che09, lan10} and hence may hint to their transitional 
nature  \citep{lan10, yan19}. No model accounts for all their properties. In Figure 1 we show what XRG look like in 
a radio map indicating both the primary and secondary jet directions.\\

In this work we identify within the gap paradigm for accreting black holes objects that potentially represent XRG. The physical reason
for their formation is the absence of frame dragging around Schwarzschild black holes, which allows for accreting clumps of cold gas 
to form  disks whose orientation differs from that of a past accretion phase. Our picture for XRG involves a jet produced in a retrograde 
disk configuration turning off and re-emerging when the spin is in the prograde direction. If this transition is bridged at the Eddington 
accretion rate, the two jets drop below observational threshold (in retrograde configuration) and re-emerge (in prograde configuration) 
within a few million years of each other so that both are visible at a single moment in time. We will show that this feature is absent 
in very powerful sources which tend to reside in clusters with largest dark matter halos, allowing us to understand both the environment 
and black hole mass distribution of XRG. In Section 2 we describe the observations and identify the jets that in the model represent XRG 
and describe their physical origin in terms of the absence of Lense-Thirring precession. In Section 3 we conclude.

\begin{figure}[ht!]
\plotone{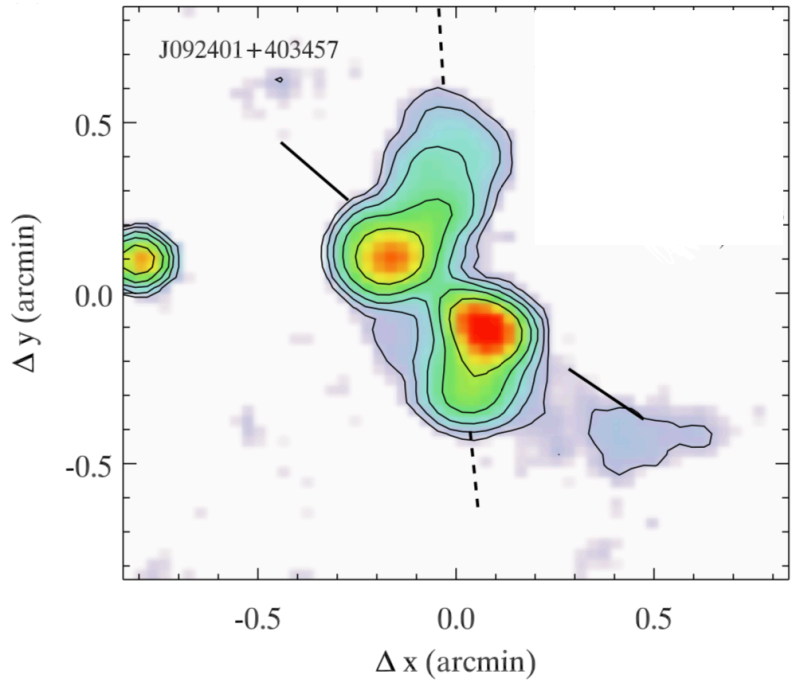}
\caption{FIRST image of X-shaped radio galaxy J092401+403457. Dotted line shows the direction of the secondary jets while the solid line shows that for the primary jets.}
\end{figure}

\section{Discussion}

\subsection{XRG – Observations}
XRG prefer isolated environments. Figure 2 shows the richness factor of XRG including that for FRII and FRI objects.  About $50\%$ of all XRG are included. 
We argue that current models provide no natural interpretation of the relative richness factor for these 3 subsets of the jetted AGN family. In the dual 
AGN model, for example, mergers are intrinsically built into the explanation which directly appears to contradict Figure 2. We will argue that XRG tend 
to avoid clusters not because they are the product of mergers but as a result of the fact that mergers in clusters tend to occur in large dark matter 
halos hosting large black holes. Backflow models also do not appear to provide an obvious way to interpret these data. In Section 2.2 we show how to understand 
Figure 2 from our time evolution framework.

\begin{figure}[ht!]
\plotone{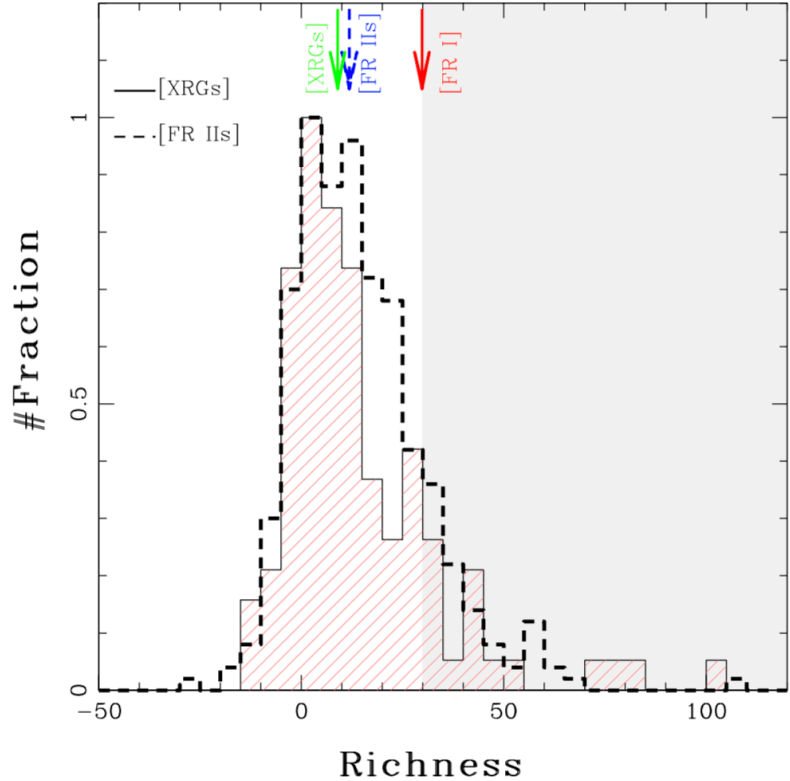}
\caption{Fraction of objects versus richness factor  (cluster environment density) for  XRG and FRII. Included are 107 XRG (out of all 225 known XRGs) and 343 
FRII from SDSS with redshift z$<$0.4 and absolute magnitude MR $<$ -19 with analysis of their environments out to 1 Mpc. XRG are shown with the solid line 
and shaded in red while FRII with the dashed line. Shown with a red arrow is the location of the richness factor for FRI. The data indicates that both XRG 
and FRII live in relatively poor environments relative to FRI. From \citet{jos19}.}
\end{figure}

XRG also appear to live near the Owen-Ledlow boundary \citep{lan10}, which divides FRII from FRI which suggests both transition and lower jet power. 
We will show that our model prescribes XRG to be objects in transition and since this occurs in the lower jet power regime, there is a coupling between 
transition and absence of the largest jet powers for XRG in the model. From the perspective of current models, there is also no clear reason why XRG 
should occur in systems with relatively smaller black hole masses, which the observations suggest (Figure 3). We will show that jets in the most massive 
black hole systems are unlikely to host an XRG, attempt to connect to the observational correlations with the optical axis of the galaxy (i.e. that the 
lobes of the primary jet align with the optical axes, \citet{cap02, sar09}), and explore the physics behind the mixed 
FRI/FRII jet morphologies and lengths of the primary and secondary jets. A crucial parameter is the spectral index which can be interpreted as resulting 
from an age difference between the two jets of order millions of years. This order of magnitude will emerge from the model. Finally, we will shed light 
on the distribution of XRG in cluster versus isolated environments, and in low excitation and high excitation radio galaxies, explain why they should 
dominate in isolated environments and high excitation state, and predict that XRG in clusters classified as low excitation are mislabeled. We will argue 
that they are best understood as transition objects in their excitation states.

\begin{figure}[ht!]
\plotone{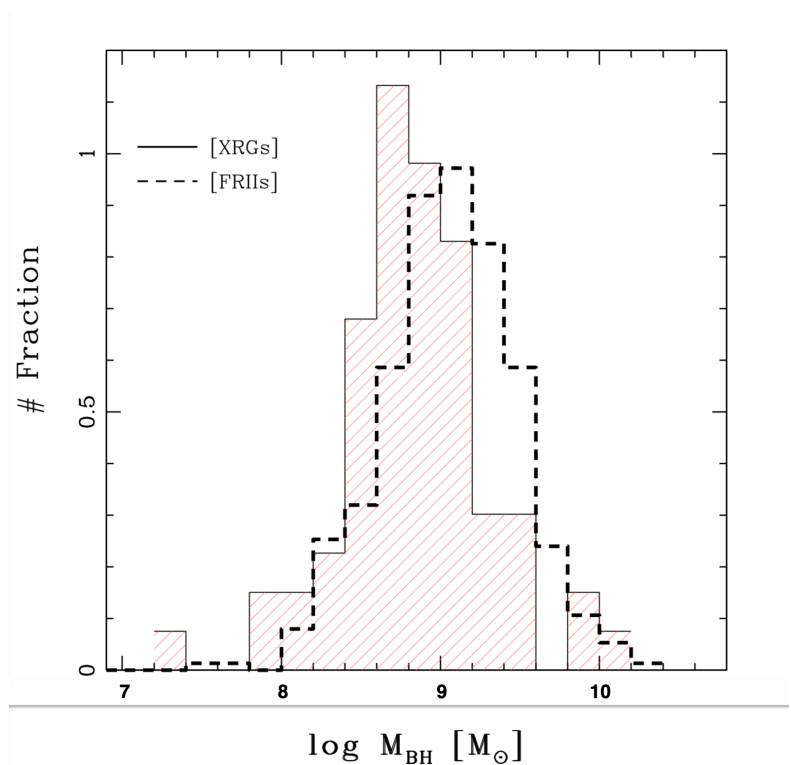}
\caption{Distribution of black hole mass in solar masses for XRG and FRII from \citet{jos19}. While the average black hole mass for FRII is 
log ($M_{BH}/M_{\odot}$) =  9.07  that for XRG is log ($M_{BH}/M_{\odot}$) = 8.81.}
\end{figure}

\subsection{XRG – theory}
Because the gap paradigm for black holes \citep{gar10} purports to explain all active galaxies in one unified framework, 
we look to its prescription for jet formation in order to find objects that may represent XRG. The main features of the model are illustrated in 
terms of three cartoons describing the time evolution of accretion around black holes that has been applied most recently to FR0 radio galaxies 
\citep{gars19}, the jet power/active lifetime correlations and environment of radio galaxies \citep{gar18}, and
the distribution of BL Lacs and flat spectrum radio quasars as a function of redshift \citep{gar18}. Figures 4, 5, and 6 represent
the subset of post-merger systems whose nuclei end up forming a retrograde accretion disk around a spinning black hole. In this work we are interested
in the evolution of accreting black holes whose jet feedback is not the most effective, which, while possible in cluster environments, is dominant in 
isolated ones (Figure 4). The crucial element in these accreting black holes is their absence of hot mode, advection dominated accretion, prior to 
their transition through zero spin. As a result of continued cold mode accretion, such systems evolve quickly from the retrograde regime to the prograde 
one and therefore rapidly transition from producing a jet of the FRII type into a jet of the FRI type. The sharp FRII/FRI division in the model is based 
on retrograde/prograde accretion but near zero spin black holes tend to be blurry in their morphological dichotomy. In Sections 2.1 and 2.2 we adopt simple
diagrams to identify the two jets that constitute the XRG and explain why zero black hole spin is crucial in potentially producing a reorientation of the primary jet.

\subsubsection{The jets of XRG}
In Figure 4 we show the evolution of cold mode accreting black holes in retrograde configuration whose mass is relatively smaller. Such systems produce weaker 
jet feedback and thus fail to influence the ISM so they have no effect on their accretion flow which persists in cold mode. Time evolution is from bottom panel 
towards top panel. If these low power FRIIs accrete near the Eddington limit, they evolve rapidly from  high retrograde spin to  high prograde spin on a timescale 
of about 100 million years. Our goal, however, is to explore the jets that are produced on opposite sides of zero spin, the one for low spin but in retrograde mode 
and the one for low spin but in prograde mode to see if there is a reasonable expectation in the model for a timescale between such jets of order a few million years. 
We wish to propose that such jets are the ones observed as XRG.\\ 

Figure 4 amounts to a more detailed version of Figure 2(a) of \citet{gar18}, in the sense that we have included low retrograde and low prograde 
spin systems as opposed to showing only the zero spin panel. For symmetry purposes Figure 4 displays -0.2 and 0.2 jets despite the fact that the -0.1 and 0.2 spin 
jets have the same power \citep{gar10}. Although the jet turns off strictly speaking only at zero spin, there is some observational threshold 
below which the jet effectively disappears. At the Eddington limit we can connect the low retrograde system with the 0.2 prograde system by estimating a timescale on 
the order of 6 million years which is compatible with observations \citep{mez11}.  In other words, the Eddington-limited timescale connecting values of spin 
that correspond to jets that are strong enough to be visible happens to be short enough that the secondary jet has not had time to dissipate. It is worth emphasizing 
that these model features are not ad-hoc, and cannot be fine-tuned to explain XRG. Note that since we are describing accretion onto relatively smaller black holes that
are found in less massive dark matter halos, the XRG described in Figure 4 are associated with isolated field environments.

\begin{figure}[ht!]
\plotone{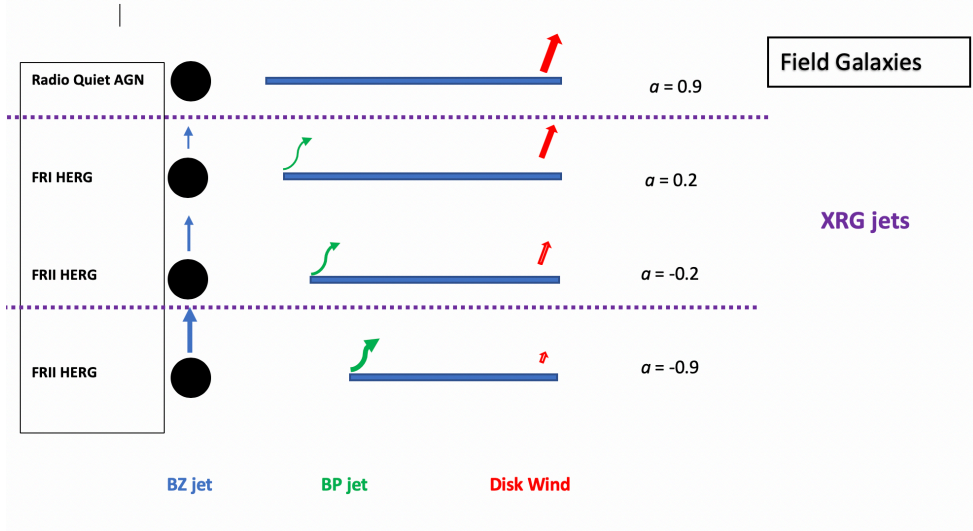}
\caption{Evolution of originally high spinning cold mode retrograde accreting black holes with smaller black hole masses and thus weaker jet feedback. Because of 
weak feedback, accretion remains in cold mode as the black hole spins down toward zero spin and then up into the prograde regime. As the spin approaches zero the jet
drops below detection threshold but will again be detected as the prograde spin is sufficiently high. Left column describes radio morphology and excitation level; 
BZ jet represented by the length and thickness of the blue arrow to characterize the Blandford-Znajek jet power; Blandford-Payne jet power from the inner disk is 
captured by the thickness of the green arrow; The red arrow captures the power of the radiative disk wind; prograde spin uses positive values while negative values
represent retrograde accretion. Details on how BZ and BP contribute to overall jet power can be found in \citet{gar10}.}
\end{figure}

In Figure 5 we show the time evolution for systems whose black holes are larger on average than those in Figure 4, and therefore produce more powerful and 
effective jet feedback that unlike for Figure 4 black holes, eventually affect the accretion flow. The uppermost panel in Figure 5 shows an advection 
dominated accretion flow (ADAF) that is the result of hot gas flowing into the inner regions. Although the details need to be worked out computationally, 
there is a basic qualitative picture that we can describe: The original FRII jet increases the entropy of the ISM and suppresses star formation such that 
before the system ends up as in the $a = 0.9$ panel, the greater galaxy is already transitioning away from the physical processes that classify it as high 
excitation, despite the fact that cold gas still fuels the black hole \citep{gar10}. We argue that these are the XRG that are 
classified as low excitation \citep{gil16} and suggest that an analysis of the distribution of emission lines will reveal weak signatures for 
the greater galaxy whereas on nuclear scales emission lines are still present and may be hidden or reprocessed by the hot halo. Note that because these 
black holes are on average more massive than in Figure 4, they live in more massive dark matter haloes, which means such XRG will distribute themselves 
in both isolated as well as in cluster environments.

\begin{figure}[ht!]
\plotone{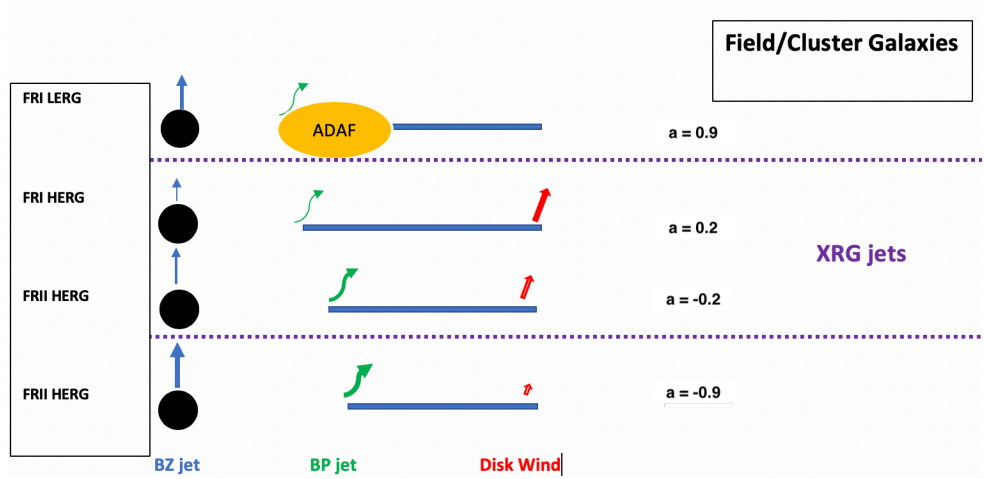}
\caption{Same as in Figure 4 but representing originally high spinning cold mode accreting black holes in retrograde configurations with higher black hole 
masses on average compared to those in Figure 4 so their jet feedback is more effective. As a result, hot gas flows into the black hole sphere of influence 
and the disk eventually transitions to an ADAF. Although the primary jet (FRI HERG) and the secondary jet (FRII HERG) share the same diagrams for their XRG 
with those of Figure 4, the greater galaxy in this case has experienced a greater jet feedback effect, such that star formation has experienced some degree 
of suppression, and the overall excitation level has dropped. Such systems can be found both in field as well as in cluster environments and may receive a 
LERG classification despite the cold mode accretion onto the black hole \citep{gil16}. 
}
\end{figure}

\begin{figure}[ht!]
\plotone{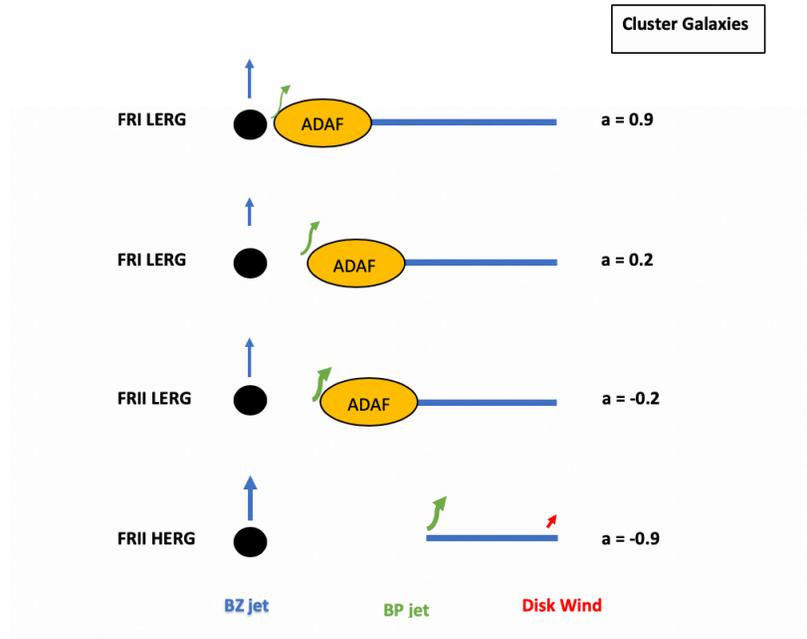}
\caption{Same as in Figures 4 and 5 but representing originally high spinning cold mode accreting black holes in retrograde configurations with highest 
black hole masses on average compared to those in Figures 4 and 5 so their jet feedback is most effective. As a result, hot gas flows into the black hole 
sphere of influence and the disk rapidly transitions to an ADAF. Because mass accretes onto the black hole at a rate at least 100 times smaller than at 
the Eddington limit (a theoretical constraint), the FRII LERG jet and the FRI LERG jet are separated in time by more than 500 million years, making both 
jets impossible to view simultaneously. Such systems dominate in cluster environments that therefore show an absence of XRG. }
\end{figure}

In places where black holes have the largest average masses, you have the largest average jet powers and largest average jet feedback on the ISM, 
a rapid transition to an ADAF and no cold gas state connecting retrograde with prograde systems (Figure 6). As a result of the ADAF phase setting 
in early while still in retrograde mode, such black holes produce FRII LERG jets at low retrograde spin and FRI LERG jets at low prograde spins 
that are separated by order a billion years. Hence, these two jets are not simultaneously visible and no XRG appear. Figures 4, 5, and 6, illustrate 
the reason why XRG tend to distribute themselves in isolated environments and less so in clusters but they also allow an understanding of why XRG that
inhabit galaxies that are becoming LERGs tend to live in clusters compared to the XRG living in galaxies that are classified as HERGs. Along the same 
lines, we can appreciate the reason why the mid-infrared signatures associated with post-merger star formation are stronger in field environments. 
The weakness of such signatures are part and parcel  of the HERG to LERG transition which increasingly comes into play as we move from Figure 4 to Figures 5 and 6.

\subsubsection{Frame dragging and jet reorientation}
In this Section we explore the physical mechanism for reorientation of the primary jet in the transition through zero black hole spin. The cold gas 
that accretes onto the black hole is supplied in clumps of mass that roughly satisfy $M_{d}/M_{BH} < 10^{-3}$ \citep{nix13} 
where $M_{d}$ is the mass of the disk and $M_{BH}$ is the black hole mass such that accretion is therefore a sequence of these clumps. Unlike \citep{kin08}, 
however, we argue that the cold gas reservoir has some average angular momentum that each clump of gas that flows toward the black hole, respects. This means 
that each sequence of clumps has angular momentum that is close to the overall average angular momentum of the cold gas reservoir, or in terms of directions satisfies

\begin{equation}
\theta_{d} - \theta_{CD} \le 180,
\end{equation}

where  $\theta_{d}$, and $\theta_{CG}$, are the angular directions of the disk and cold gas torus angular momenta, respectively. The bottom line is that chaotic 
accretion does not occur. When the black hole spins rapidly, the difference in the directions of the angular momenta of the black hole and disk are subject to 
frame dragging and the Lense-Thirring effect which brings the angular momenta into alignment or counter-alignment \citep{kin05, kin08}, of which we are 
interested in the latter. Counter-alignment between disk and black hole is stable under restrictive conditions on black hole and disk mass which makes such 
systems a minority among the total active galaxy population \citep{garc19}. This picture amounts to a continuous accretion scenario 
whereby an initially retrograde accreting black hole continues to spin down toward zero spin and eventually up into the prograde direction. However, Lense-Thirring 
precession depends on the presence of frame dragging which requires a spinning black hole. At zeroth order frame dragging imparts an angular velocity that is proportional 
to black hole spin via

\begin{equation}
\omega = 2GM_{c}a/c^{2}r^{3}
\end{equation}

where $G$ is the gravitational constant, $c$ is the speed of light, $a$ is the dimensionless spin of the black hole, and $r$ is the radial distance from the black hole.  
Hence, the impetus for alignment or counter-alignment of the angular momenta progressively vanishes as the black hole spins down toward zero spin. Therefore, the clumps
feeding the black hole near zero spin form thin disks that tend to retain their original $\theta_{CG}$ values, the clumps illustrated further out in Figures 7 and 8.  
We show the details of this in Figures 7 and 8 with a spinning black hole in Figure 7 and a Schwarzschild black hole in Figure 8. In the former we see the Bardeen-Petterson 
effect \citep{bar75} aligning (or anti-aligning) the angular momenta near the black hole, whereas in the latter we see the inner disk forming with 
the same angular momentum direction as that of the inflowing clumps.

\begin{figure}[ht!]
\plotone{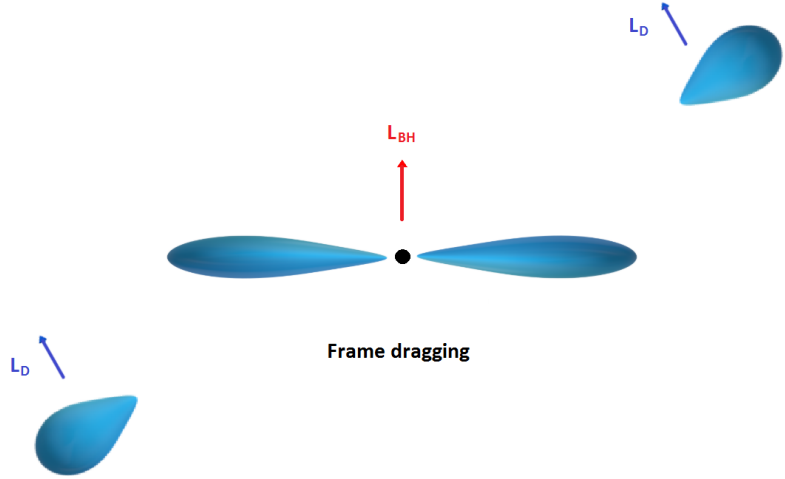}
\caption{Illustration of the Bardeen-Petterson effect on the inflowing cold gas. The incoming gas ends up in the plane determined by black hole spin direction.}
\end{figure}

\begin{figure}[ht!]
\plotone{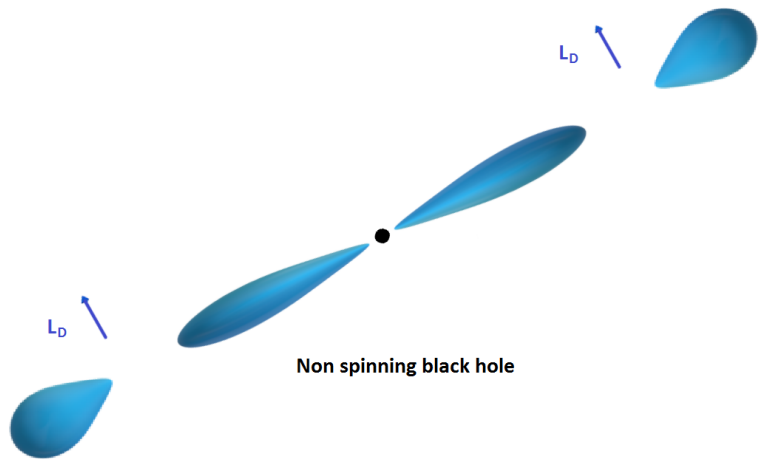}
\caption{Illustration of the absence of the Bardeen-Petterson effect on the inflowing gas. The accretion disk retains the direction of its original angular momentum about the black hole.}
\end{figure}

As a result of the absence of Bardeen-Petterson for zero spin black holes, a new plane of accretion is generated which produces a new jet direction that becomes visible 
after a few million years when the spin is about $0.2$ and the observational threshold for the jet is crossed.\\ 

A number of features drop out of this model. First, notice that because the retrograde jet is more powerful than the prograde jet for a given value of spin 
(in Figures 4 and 5 the choice is an absolute spin value of $0.2$), the earlier jet is more powerful and collimated, hence travels further out into the ISM 
than the latter prograde jet. This qualitatively explains the observation that the secondary jet extends further than the primary jet in XRG. However, this 
is not a dominant effect since the spin is relatively low and environment may have a comparably larger effect. In other words, near zero spin, a denser 
environment can turn an engine generated FRII-like jet into an FRI-like jet. Note that these ideas form the backdrop for the explanation of the Owen-Ledlow diagram 
\citep[see][Section 3.4]{gar10}.\\ 

Second, the possibility of viewing the retrograde and prograde jets at the same time is only possible because cold mode accretion allows for an evolution from one 
jet to the other in just a few million years. If ADAF accretion were to replace the thin disk during the transition through zero spin, the timescale between the 
retrograde and prograde jets would increase dramatically (by at least a factor of $100$) and no two jets would be observed. This is due to the fact that the accretion
rate in ADAF mode compared to that at the Eddington limit has been theoretically determined to be $(dM/dt)_{ADAF} < 0.01(dM/dt)_{Edd}$. This can be seen in Figure 6 where
the ADAF disk sets in early when the black hole is still accreting in retrograde mode. Hence, there is an explanation for why XRG emerge in systems near the FRII/FRI 
boundary of the Owen-Ledlow diagram and not in systems that rapidly transition to ADAFs such as those that are connected to the recently identified class of FR0 radio 
galaxies \citep{gars19}. The jet power, in short, must either not be so large that it affects the accretion flow or, it must not affect the accretion state 
rapidly, allowing the system to transition well into the prograde regime before it becomes an ADAF. Notice that indeed the black hole masses in XRG appear to be slightly
smaller \citep{jos19}. From the model viewpoint, radio morphology gets blurry near zero spin and environment will tend to wash out engine-based signatures. 
To the extent, therefore, that the secondary is of the FRII type, the Blandford-Znajek and Blandford-Payne jets that give rise to power and collimation are weaker 
at low retrograde spin than at high retrograde spin so there should be a difference in these features from a model perspective.\\ 

Third, XRG emerge in systems with weak jet feedback in the model which means they tend to be dominant in isolated environments as shown in Figures 4 and 5 although they 
can exist in clusters as well (Figure 5). The model prediction for XRG that form in field environments and that continue to accrete to high prograde spin values is 
a radio quiet quasar/radio quiet AGN as shown in the uppermost panel in Figure 4. For those of Figure 5, instead, continued accretion generates FRI radio galaxies. 
Note that in addition to the recent transition through zero spin for Figure 5 objects, there is an evolution in the state of accretion that is not present in Figure 4. 
Jet feedback in Figure 5 objects has slowly heated the ISM and shutdown star formation and the hot gas that now permeates the greater galaxy will eventually flow back 
into the black hole sphere of influence as an ADAF (Figure 5 upper panel). But the primary jet of the XRG is formed when the spin is still low (Figure 5 for a = 0.2 spin). 
Hence, although the greater galaxy is transitioning away from the processes associated with strong emission lines, the near black hole disk is still cold. The model 
therefore prescribes that XRG classified as low excitation radio galaxies (LEGs) \citep{gil16}, are actually transition objects that until recently
still produced strong emission lines. Interestingly, some of the LEG classified XRG are close to the HEG/LEG boundary \citep{gil16}.\\ 

Fourth, the model predicts that the merger that produced the original FRII HERG (Figures 4 and 5) occurred millions of years in the past of the XRG and that XRG are 
therefore not directly related to mergers, which seems compatible with observations \citep{lan10}. The model also allows one to recast an old question in a 
new context \citep{san87}: what is the connection between the directions of the XRG jets and the optical axes of the galaxy? There are two directions that 
emerge from the model: The original direction of the black hole angular momentum and the direction of the angular momentum of the molecular torus that feeds the black hole. 
These are the directions for the jets in XRG. Hence, the question of how the jet directions are related to minor and major optical axes in the model reduces to understanding 
the connection between original black hole angular momentum and the angular momentum of the torus that feeds the black hole and the optical axes.\\ 

And, finally, we point out that in the model the primary jet of the XRG is an FRI quasar in its youth \citep{blu01, kim16}. If accretion 
continues in such objects, the system evolves either into a radio quiet quasar/AGN (Figure 4 upper panel) or into an FRI radio galaxy (Figure 5 upper panel) 
which explains the dearth of FRI quasars. The answer to the longstanding question of where the FRI quasars are appears here: They form only in a narrow range of spin 
and tend to come in the form of XRG. In closing, we point out that some XRG have been identified with a different origin. Our model cannot account for such objects 
as the nucleus in our model is fixed. However, we can predict that such XRG would not be constrained in the same way as the XRG discussed in this work. For example, 
there would be no reason for low black hole spins and no preference for isolated environments for such objects.

\section{Conclusions}
Exploring two of three ways black holes evolve in the gap paradigm, we have identified a space for XRG as transition objects between retrograde and prograde accreting
black holes and therefore as objects with low spin. If black hole spins down at near Eddington accretion rates, the transition through zero spin allows for the jet 
to effectively disappear observationally and to re-emerge when the black hole has spun up sufficiently in the prograde direction. We have shown that a few million years 
is a reasonable timescale between jets, creating conditions that allow for both jets to be visible. Because frame dragging disappears at zero spin, an opportunity emerges 
in the model to invoke the formation of a new plane of accretion that is determined by the angular momentum of the gas reservoir, setting the stage for the two jets 
to experience different orientations. Our model accounts for the weakness of the primary jet compared to the secondary, small differences in spectral index for the jets, 
the preference of XRG in isolated field environments over clusters, the reason XRG tend to live near the Owen-Ledlow boundary line, low spinning black holes for both primary 
and secondary jets and lower black hole masses compared to other jetted AGN, their connection to FRI quasars, their excitation classification and their relation to environment, 
and their future states as either radio quiet quasar/AGN or FRI radio galaxies if fuel continues to be available.

\acknowledgements
We thank the anonymous referee for emphasizing clarity in crucial places. DG thanks Alessandro Capetti and Ranieri D. Baldi for helpful discussion.

\end{document}